\documentclass[useAMS,usenatbib]{mn2e}
\usepackage{graphicx}
\usepackage{times}

\def\farcs{\hbox{$.\!\!^{\prime\prime}$}}

\title[Spectroscopic orbits of nearby dwarfs]{Spectroscopic orbits of nearby  solar-type dwarfs. II.}

\author[Gorynya \& Tokovinin]{N.~A.~Gorynya$^{1,2}$\thanks{E-mail: gorynya@sai.msu.ru} 
 \& A.~Tokovinin$^{3}$\thanks{E-mail: atokovinin@ctio.noao.edu} \\
$^1$Institute of Astronomy,  Russian Academy of Sciences, 48 Pyatnitskaya Str,
   Moscow, 109017,  Russia \\
$^2$ Lomonosov Moscow State University, Sternberg State Astronomical
Institute,  13 Universitetskij prospekt, Moscow, 119991, Russia \\
$^3$Cerro Tololo Inter-American Observatory, Casilla 603, La Serena, Chile\\
}

\begin{document}

\date{-}

\pagerange{\pageref{firstpage}--\pageref{lastpage}} \pubyear{2018}

\maketitle

\label{firstpage}

\begin{abstract}
Several nearby  solar-type dwarfs  with variable radial  velocity were
monitored to find their  spectroscopic orbits.  First orbital elements
of 15 binaries  (HIP 12144, 17895, 27970, 32329,  38636, 39072, 40479,
43004,  73700, 79234,  84696, 92140,  88656, 104514,  and  112222) are
determined.  The  previously known orbits  of HIP 5276,  21443, 28678,
and 41214 are confirmed and updated.  The orbital periods range from 2
days to  4 years.   There are 8  hierarchical systems  with additional
distant companions  among those 19  stars.  The outer visual  orbit of
the  triple system  HIP~17895 is  updated and  the masses  of  all its
components are estimated.  We  provide radial velocities of another 16
Hipparcos  stars without orbital  solutions, some  of those  with long
periods or false claims of variability. 
\end{abstract}


%%--------------------------------------------------------------
\section{Introduction}
\label{sec:intro}

This  paper  continues the  work  on  spectroscopic  orbits of  nearby
solar-type    stars   introduced    in    our   previous    publication
\citep[][hereafter GT14]{GT14}.   Stars with variable  radial velocity
(RV) and unknown orbits that  belong to the 67-pc sample \citep{FG67a}
were selected  for monitoring.   Many of those  spectroscopic binaries
(SBs)   were   discovered  by   the   Geneva-Copenhagen  Survey,   GCS
\citep{N04}.  Their  unknown periods and mass  ratios prevent detailed
statistical study  of this  important volume-limited sample.   A large
number  of such  binaries were  followed by  D.~Latham at  Center for
Astrophysics, leading to hundreds  of orbital solutions (D.~Latham, in
preparation).  Still, not all spectroscopic binaries are covered.

As  in  GT14, our  aim  is  to complement  the  existing  work and  to
determine,  whenever  possible,  spectroscopic  orbits.   The  objects
covered in this study are listed in Table~\ref{tab:list}, where visual
magnitudes  and spectral  types are  taken from  SIMBAD, trigonometric
parallaxes $p$ from  the {\it  Hipparcos-2}  \citep{HIP2}  or {\it  Gaia}
\citep{Gaia}  catalogues (the latter  are marked  by asterisks), and
the masses are estimated from  the absolute magnitudes \citep{FG67a}.  All
stars are bright; their spectral types range from F3V to G8V. The last
column resumes  our results  by indicating the  orbital periods  or RV
variability; astrometric  acceleration detected by  {\it Hipparcos} is
noted as well.

\begin{table}
\centering
\caption{List of 35 observed stars}
\label{tab:list}
\medskip
\begin{tabular}{rr c c c c l } 
\hline
HIP & HD & $V$ & Spectral & $p$ & $M_1$ & Note \\
    &    & mag & type     & mas            & $M_\odot$ & \\  
\hline
 5276 & 6611 & 7.24   & F5   & 19.4 & 1.29 & SB1, 74d \\
12444 & 16673 & 5.79 & F8V  & 46.0 & 1.20 & SB1, 37d \\ 
17895 & 24031 & 7.23 & F8V  & 19.6 & 1.14 & SB3, 251d \\ 
20693 & 28069 & 7.36 & F7V  & 21.3*& 1.26 & Const.     \\
21443 & 28907 & 8.61 & G5   & 15.5 & 1.11 & SB1, 2.1d \\
24076 & 33507 & 7.42 & F6V  & 17.4 & 1.14 & Const. \\
26444 & 37271 & 7.64 & F5   & 20.4*& 1.20 & Const.   \\
27878 & 39570 & 7.76 & G2   & 18.7 & 1.20 & Const. \\
27970 & 39899 & 7.73 & F5V  & 16.0 & 1.27 & SB1, 15d \\
28678 & 41255 & 7.47 & F9V  & 15.8 & 1.21 & SB2, 148d \\
31480 & 46871 & 7.51 & G0V  & 14.0*& 1.35 & Trend, acc. \\
32329 & 48565 & 7.18 & F8   & 20.3 & 1.28 & SB1, 73d \\
36836 & 60552 & 6.71 & F7II & 22.5 & 1.34 & Trend, acc. \\
38636 & 64273 & 8.36 & G5   & 16.5 & 1.12 & SB1, 66d \\
39072 & 65156 & 8.67 & G0   & 15.6 & 1.09 & SB1, 22d \\
40479 & 69151 & 8.16 & F6V  & 15.9 & 1.18 & SB1, 58d  \\
41214 & 70937 & 6.03 & F2V  & 15.8 & 1.61 & SB2, 28d \\
43004 & 74655 & 7.59 & F8   & 18.4 & 1.24 & SB1, 70d \\
43393 & 75530 & 9.18 & G8V  & 20.0*& 0.94 & Var., slow \\
48391 & 85238 & 7.80 & G0   & 22.0 & 1.11 & Const.? \\
55982 & 99739 & 7.24 & F6V  & 20.4*& 1.29 & Var. 35d? \\
56282 & 100269& 8.09 & F8V  & 15.5*& 1.19 & Var. 122d? \\ 
66290 & 118244& 6.99 & F5V  & 21.6 & 1.31 & Trend, acc. \\ 
69238 & 124086& 8.29 & G0   & 15.2 & 1.18 & Trend, acc. \\
73700 & 133460& 7.28 & F8V  & 15.2 & 1.42 & SB1, 200d \\ 
73765 & 133725& 7.49 & F8   & 19.5*& 1.21 & Var., 1.5d? \\
79234 & 145605& 7.72 & F5   & 18.3 & 1.21 & SB1, 9d \\ 
81312 & 149890& 7.10 & F8V  & 26.0*& 1.19 & Const. \\
84696 & 156635& 6.66 & F7V  & 24.8 & 1.30 & SB1, 3.6yr \\
85963 & 159307& 7.40 & F3V  & 15.3 & 1.37 & Var., acc. \\
88656& 165360& 7.09 & F8V  & 18.3* & 1.25 & SB2, 4.3d \\ 
92140 & 173614& 7.00 & F5V  & 18.4 & 1.31 & SB1, 588d \\
104514& 201639& 8.14 & F8   & 15.5 & 1.20 & SB1, 266d \\
109122& 209767& 7.12 & F3V  & 16.9 & 1.40 & Var., 1540d? \\
112222& 215243& 6.51 & G8IV & 23.6 & 1.37 & SB1, 4.5yr \\
\hline
\end{tabular}
\flushleft{{\it Note:}  See Section~1 for the description of the columns.}
\end{table}

%%--------------------------------------------------------------
\section{Observations}
\label{sec:obs}

The  observations were started  in 2012  at the  1-m telescope  of the
Crimean Astrophysical  Observatory sited in Simeiz,  Crimea.  The last
observing  season  used  in  this  work was  in  October  2017.   Radial
velocities were measured by the CORAVEL-type echelle spectrometer, the
Radial  Velocity Meter (RVM).   This instrument  is based  on analogue
correlation of  spectrum with a physical mask,  where slits correspond
to the  spectral lines  \citep{RVM}. The RVs  are measured  by fitting
Gaussian curves to the observed  correlation dips, with the velocity zero point
determined from observations of  RV standards.  Further information on
the  observing  method and  its  limitations  can  be found  e.g.   in
\citep{TG01}.  The  RV precision  reaches 0.3\,km~s$^{-1}$, but  it is
worse  for  stars with  shallow  correlation  dips  and/or fast  axial
rotation.  

For some  stars, additional RVs  were measured with the  fiber echelle
and  the  CHIRON  spectrometers   \citep{CHIRON}  at  the  CTIO  1.5-m
telescope. The reader should  refer to \citep{LCO,CHI} for information
on those instruments and the  data reduction methods.  The RVs are
  derived from  the fiber echelle  and CHIRON spectra  by cross-correlation with
  the mask  based on the solar  spectrum and therefore  do not require
  zero-point adjustment. Indeed, we see no systematic difference with
  the RVM velocities.

%%--------------------------------------------------------------
\section{New orbits}
\label{sec:orb}

\begin{figure*}
\centerline{
\includegraphics[width=17cm]{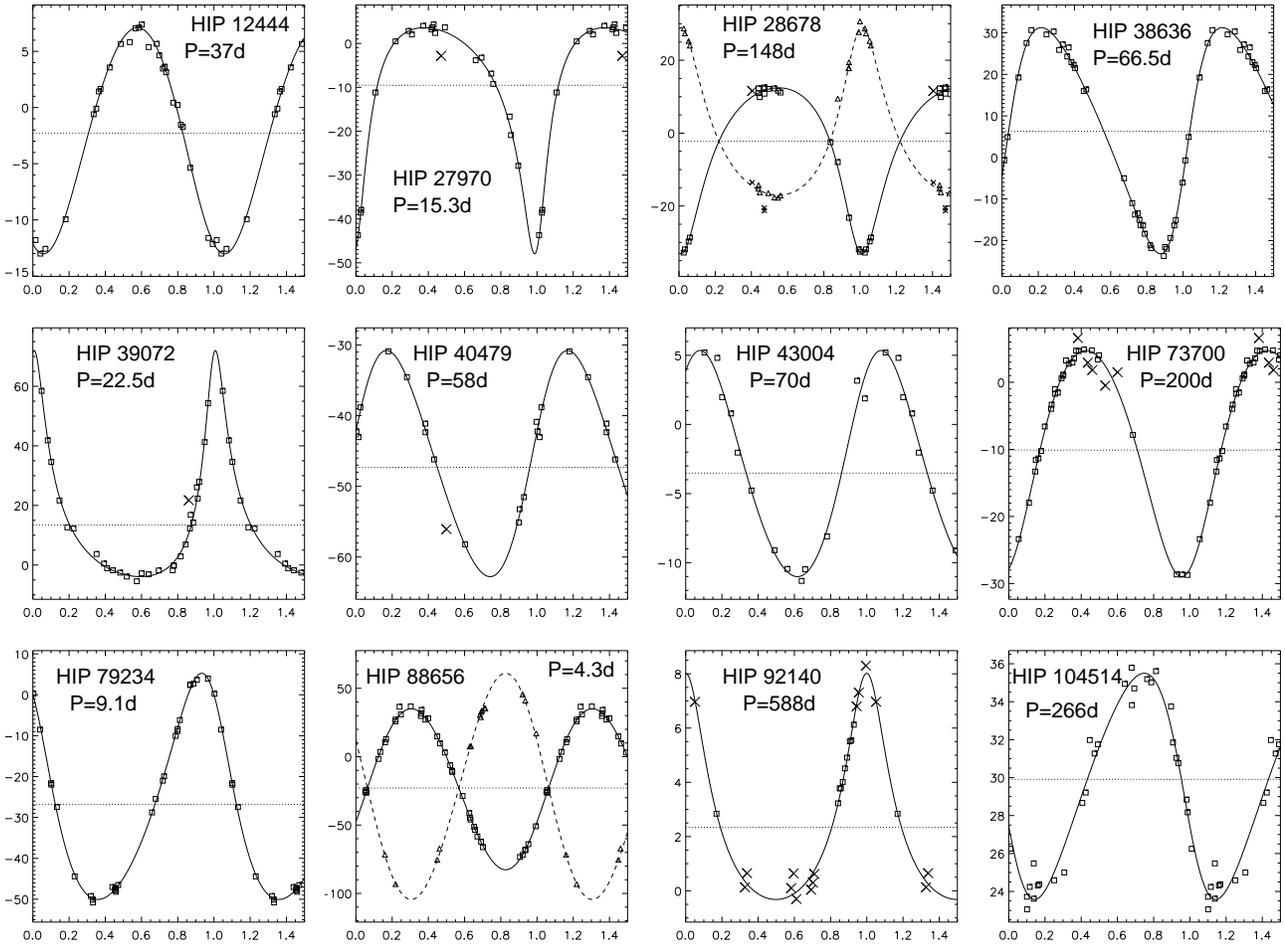}
}
\caption{Radial velocity curves. Each panel plots the orbital phase on
  the  horizontal axis  and the  RV (in  km~s$^{-1}$) on  the vertical
  axis.  The measurements are plotted as squares and triangles for the
  primary and secondary components, respectively.  The full and dashed
  lines mark the RV curves,  the horizontal dotted line corresponds to
  the centre-of-mass velocity. The crosses mark less accurate RVs that
  were given a low weight.
\label{fig:orb1} }
\end{figure*}

\begin{figure*}
\centerline{
\includegraphics[width=17cm]{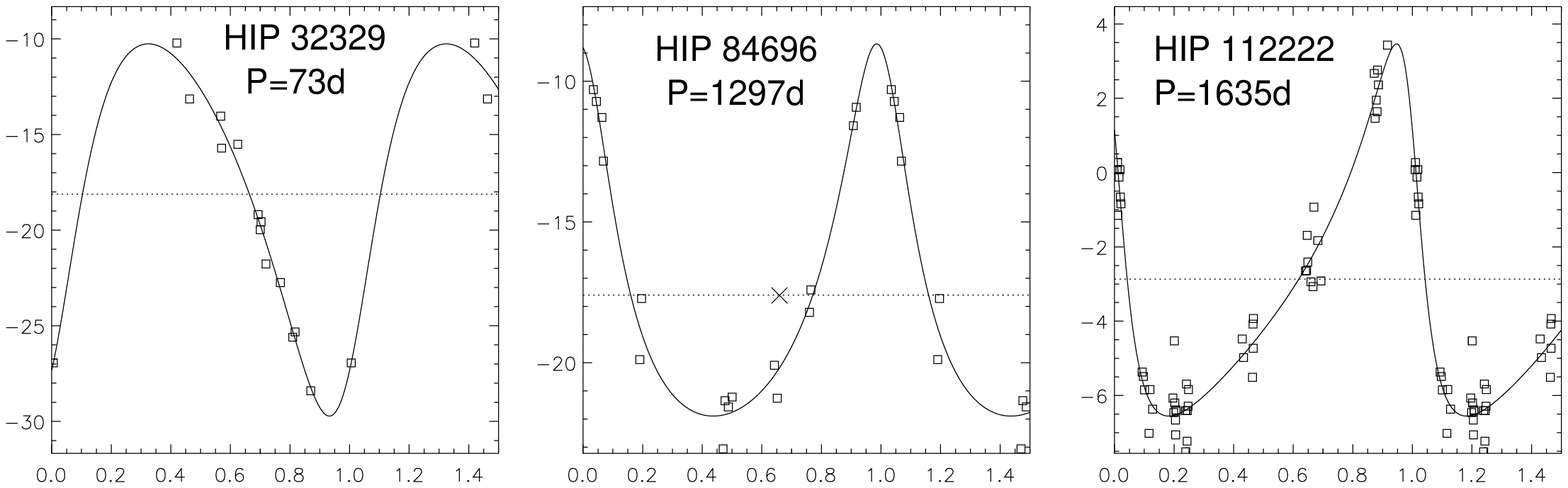}
}
\caption{Radial velocity curves for tentative orbital solutions.
The crosses mark  RVs that were given low weights.
\label{fig:orb2} }
\end{figure*}

Identifications and  basic parameters of stars  with orbital solutions
are  given in  Table~\ref{tab:list}.  The  orbital elements  and their
errors are  listed in  Table~\ref{tab:orb} in standard  notation.  Its
last columns contain the total  number $N$ of the RV measurements, the
rms residuals to the orbit,  and the mass estimates.  For double-lined
binaries  we  provide $M  \sin^3  i$,  for  single-lined binaries  the
minimum secondary mass $M_{\rm min}$ is listed, computed from the mass
function and the estimated primary  mass $M_1$ by solving the equation
$M_{\rm  min} =  4.695\,10^{-3}\, K_1  P^{1/3}(1 -  e^2)^{0.5}  (M_1 +
M_{\rm min})^{2/3}$.  The RV curves are plotted in Fig.~\ref{fig:orb1}
and  Fig.~\ref{fig:orb2}.  We  do not  provide plots  for  the updated
orbits  from GT14.   In  the orbital  fits,  the RVs  are weighted  as
$1/(\sigma_i^2  +  0.3^2)$, where  $\sigma_i$  are  the individual  RV
errors determined by  the Gaussian fits of the  correlation dips, with
the instrumental  error of  0.3\,km~s$^{-1}$ added in  quadrature. The
RVs derived from the  unresolved blended dips of double-lined binaries
are given a very low weight by artificially  increasing the errors
  to  $>20$ km~s$^{-1}$.    Otherwise, the double dips are fitted by
two Gaussian curves.   The formal errors are also increased  by 1 or 2
km~s$^{-1}$ for stars with fast rotation and shallow correlation dips.
 Short-period  binaries are expected to have  circular orbits. For
  HIP~21443 ($P =  2$ d) we fix the elements  $e=0$ and $\omega=0$ and
  fit  the remaining  four  elements $P,T,K_1,  \gamma$. However,  the
  small  eccentricity  of  HIP~88656  ($P= 4.3$  d)  is  significantly
  different from zero,  hence the eccentric orbit is  retained.   The
individual  RVs  and residuals  to  the  orbits  are provided  in  the
Supplementary material (Table~4).

\begin{table*}
\centering
\caption{Orbital elements}
\label{tab:orb}
\medskip
\begin{tabular}{l ccc c ccc c c c }
%\hline
\hline
HIP & $P$ & $T_0$ & $e$ & $\omega$ & $K_1$ & $K_2$ & $\gamma$ & $N$ & rms & Mass$^{\rm a}$ \\
    & days & +2400000 &  & deg     & km~s$^{-1}$ & km~s$^{-1}$ & km~s$^{-1}$      &     & km~s$^{-1}$ & $M_\odot$ \\   
\hline
 5276 & 74.1607 & 56278.110         & 0.289      & 146.2 & 25.12  & \ldots& 9.85   & 71 & 0.69 & 0.77 \\
         & $\pm$0.0020 & $\pm$0.146 & $\pm$0.004 & $\pm$0.6 & $\pm$0.125 & \ldots& $\pm$0.07  & \ldots   & \ldots     & \ldots\\
 12444 & 37.088  & 56618.21          & 0.074      & 155.7 & 10.07  & \ldots& $-$2.28   & 27 & 0.38 & 0.20 \\
         & $\pm$0.0067 & $\pm$0.93   & $\pm$0.012 & $\pm$8.7  & $\pm$0.14  & \ldots& $\pm$0.08 & \ldots& \ldots&\ldots \\
  17895B & 250.82 & 57255.97          & 0.420      & 344.9    & 20.45   & 23.90 & 7.35            & 32 & 0.55 & 0.91 \\
         & $\pm$0.16 & $\pm$0.68    & $\pm$0.006  & $\pm$1.3 & $\pm$0.19 & $\pm$0.20 & $\pm$0.09 & 19 & 0.54 & 0.78 \\ 
21443 & 2.06198 & 56604.5096        & 0.000      & 0.0 & 23.73    &\ldots& $-$5.26  & 21 & 1.14  & 0.17 \\
         & $\pm$0.0001 & $\pm$0.0022 & fixed & fixed   & $\pm$0.15 & \ldots& $\pm$0.11 & \ldots& \ldots& \ldots\\
 27970 & 15.3230 & 56600.957         & 0.513     & 197.0 & 25.79  & \ldots    & $-$9.54  & 21 &  0.82 & 0.36 \\
         & $\pm$0.0005 & $\pm$0.030  & $\pm$0.006& $\pm$0.9& $\pm$0.36& \ldots& $\pm$0.11 &  \ldots  &\ldots &\ldots \\
 28678 & 148.43   & 56603.67         & 0.361      & 170.8 & 22.78  & 22.98  & $-$2.23             & 20 & 0.64 & 0.60 \\
         & $\pm$0.05   & $\pm$0.62   & $\pm$0.006 & $\pm$2.7 & $\pm$0.15  & $\pm$0.14  & $\pm$0.09 & 20 & 0.78 & 0.59 \\
 32329 &  73.356  & 56919.04         & 0.255      & 221.0 &  9.73 &  \ldots  & $-$18.12   & 14   & 0.67 & 0.24 \\
         & $\pm$0.034  & $\pm$1.27   & $\pm$0.098  & $\pm$10.7 & $\pm$1.13  & \ldots     & $\pm$0.68 &  \ldots  &\ldots  & \ldots \\
 38636 & 66.500  & 57143.82          & 0.266      & 251.7 & 27.23  & \ldots& 6.30   & 33 & 0.76 & 0.76 \\
         & $\pm$0.006  & $\pm$0.16   & $\pm$0.003 & $\pm$1.1 & $\pm$0.11  & \ldots& $\pm$0.08 &  \ldots & \ldots &  \ldots \\
 39072 & 22.5139 & 57329.781         & 0.555      & 348.2 & 37.97  & \ldots& 13.43  & 30 & 1.48 & 0.59 \\
         & $\pm$0.0004 & $\pm$0.017  & $\pm$0.005 & $\pm$0.4 &  $\pm$0.38   & \ldots& $\pm$0.10 &  \ldots& \ldots  & \ldots \\
 40479 & 58.206  & 56902.87          & 0.118      & 286.5 & 16.00  & \ldots& $-$47.31  & 14 & 0.80 & 0.39 \\
         & $\pm$0.010  & $\pm$1.20   & $\pm$0.013 & $\pm$7.8 & $\pm$0.28   & \ldots& $\pm$0.20 & \ldots & \ldots & \ldots \\
 41214 & 27.8890 & 56230.73          & 0.331      & 162.9 & 44.37  & 57.65 & $-$28.26             & 24 & 0.96 & 1.46 \\
         & $\pm$0.0013 & $\pm$0.17   & $\pm$0.008 & $\pm$1.6 & $\pm$1.00   & $\pm$0.78  & $\pm$0.42 &17 & 0.89 & 1.12 \\
 43004 & 70.69   & 57602.08          & 0.103      & 325.1 & 8.18   & \ldots& $-$3.52   & 13 & 0.79 & 0.20 \\
         & $\pm$0.06   & $\pm$2.81   & $\pm$0.023 & $\pm$13.9& $\pm$0.19  & \ldots& $\pm$0.15 &  \ldots&  \ldots& \ldots \\
 73700 & 200.23  & 57207.92          & 0.127       & 201.0 & 16.86 & \ldots& $-$10.11 & 34 & 0.52 & 0.78 \\
         & $\pm$0.18   & $\pm$2.92   & $\pm$0.010 & $\pm$5.1 & $\pm$0.11  & \ldots& $\pm$0.16 & \ldots & \ldots & \ldots\\
 79234 & 9.0752 & 57245.949          & 0.196 & 35.8 & 27.66  & \ldots     & $-$26.79  & 28 & 0.66 & 0.36 \\
         & $\pm$0.0002 & $\pm$0.029  & $\pm$0.004 & $\pm$1.3 & $\pm$0.12  &\ldots & $\pm$0.08 &  \ldots&  \ldots& \ldots\\
   84696 & 1297      & 56652.7       & 0.358      & 11.7     & 6.61   &\ldots & $-$17.60  & 17  & 0.65 &  0.46 \\
         & $\pm$16      & $\pm$12.2  & $\pm$0.032 & $\pm$3.7 & $\pm$0.42  &\ldots & $\pm$0.13 &  \ldots & \ldots & \ldots\\
 88656 & 4.26928  & 57286.716        & 0.037      &246.2      &  58.83  & 82.70      & $-$22.95  &  40 & 0.92 & 0.73 \\
         & $\pm$0.00001& $\pm$0.043  & $\pm$0.001 & $\pm$3.6  & $\pm$0.17  & $\pm$0.28 & $\pm$0.08 & 15 & 0.80 & 0.52 \\
 92140 & 587.76   & 55532.65         & 0.362      &  0.4      & 4.17      & \ldots     & 2.34 &  22 & 0.25 & 0.20 \\
         & $\pm$2.88   & $\pm$9.17   & $\pm$0.064 & $\pm$6.2  &  $\pm$0.21  & \ldots    & $\pm$0.15 &  \ldots& \ldots & \ldots \\
 104514 & 265.81   & 57258.68        & 0.177     &110.6       & 5.96       & \ldots       & 29.91  & 28 & 0.69 & 0.22 \\
         & $\pm$0.54   & $\pm$4.40   & $\pm$0.022 & $\pm$6.4  & $\pm$0.12  & \ldots& $\pm$0.09 &  \ldots& \ldots &  \ldots\\
   112222 & 1635      & 56207.2      & 0.492      & 57.5      & 5.01    &\ldots  & $-$2.87 & 51  & 0.52 & 0.35  \\
         & $\pm$20      & $\pm$6.1    & $\pm$0.040 & $\pm$2.3 &   $\pm$0.36  &\ldots & $\pm$0.09 &  \ldots & \ldots & \ldots \\
\hline
%\hline
\end{tabular}
\flushleft{ {\it Note:}  $^{\rm a}$   $M
  \sin^3 i$ for double-lined binaries and $M_{\rm min}$ for single-lined binaries.   }
\end{table*}

We now  comment on each spectroscopic binary  individually, except the
triple-lined  system  HIP~17895 which  is  deferred  to the  following
Section.  The masses of the  primary components are derived from their
apparent  $V$  magnitudes  and  trigonometric  parallaxes  using  
  standard   relations  for   main   sequence  dwarfs   \citep{FG67a};
  corrections are made for  the magnitudes of double-lined binaries. 
The existence of additional (tertiary) visual components is noted.

{\bf HIP 5276} was discussed  in GT14. Its preliminary 74-day orbit is
now  definitive.  This is  a  triple  system  with the  physical  visual
companion B at 6\farcs2 \citep{Roberts2015}. 

{\bf  HIP  12444}  has  now  a  well-defined  37-day  nearly  circular
orbit. The minimum  secondary mass is 0.2 ${\cal  M}_\odot$. This star,
located  at   21.7\,pc,  has  attracted   considerable  attention  (194
references  in   SIMBAD),  and  it   is  surprising  that   since  its
identification as a spectroscopic binary  in the GCS no orbit has been
determined    despite   spectroscopic    surveys    like   those    of
\citet{Fuhrmann2017}.

{\bf HIP 21443}  with a 2-day period was presented  in GT14. The orbit
is now  strengthened by additional  measurements. This is a  triple system
with the physical companion at 5\farcs7 discovered by \citet{RoboAO}. 

{\bf  HIP 27970}  is  a  {\it Hipparcos}  acceleration  binary with  a
variable  RV detected  by the  GCS. The  15-day orbit  determined here
implies that the system is  triple, as the spectroscopic period is too
short to cause any detectable acceleration; the period of the tertiary
companion remains unknown.  The astrometric binary was not resolved by
speckle   interferometry  at   the  Southern   Astrophysical  Research
Telescope,  SOAR \citep[][and  references  therein]{SOAR}, placing  an
upper limit on its  separation and mass.  The centre-of-mass velocity,
$-9.5$  km~s$^{-1}$, is  very close  to  the mean  velocity of  $-9.6$
km~s$^{-1}$ given in the GCS (no long-term RV trend).

{\bf HIP 28678} is a  double-lined binary with nearly equal components
for which a  tentative 163-day orbit was proposed  in GT14. The period
is  now  revised to  148  days,  and the  new  RV  curve  is shown  in
Fig.~\ref{fig:orb1}.  The semimajor axis is 11\,mas. The spectroscopic
masses are two times less than the masses estimated from the luminosity,
hence the orbital inclination should be around 53$^\circ$. The binary
can be resolved by speckle interferometry at 8-m telescopes.

{\bf HIP  32329} has a preliminary  orbit with $P=73$  days. A similar
period has been determined by D.~Latham (2015, private comminucation),
lending independent verification of  our tentative orbit. Latham noted
that the system  is triple, and, indeed, it  is an acceleration binary
in {\it  Hipparcos}. The outer period  is not known  and the companion
has not  been resolved by speckle interferometry  and adaptive optics.
This  is   a  barium  star  \citep{Allen2006},   suggesting  that  the
spectroscopic companion is most likely a white dwarf; its minimum mass
is 0.25  ${\cal M}_\odot$.  The  relatively large eccentricity  of the
inner orbit,  $e=0.26$, is  at odds with  the suggested nature  of the
companion  (a  nearly  circular   orbit  is  expected).  However,  the
eccentricity could be pumped up by the tertiary companion.

{\bf  HIP 38636}  has a  variable  RV according  to the  CGS. Now  its
66.5-day orbit  is determined. This is  a triple system  with a common
proper motion (CPM) companion at 98$''$.

{\bf HIP 39072},  like the previous object, is  a spectroscopic binary
detected by the GCS which now  has a 22.5-day orbit.  No visual
components are known.

{\bf HIP 40479} has a  58-day orbit.  The visual companion
at  31\farcs2 noted by  \citet{RoboAO} is  likely optical  because its
position in  2MASS is not the  same.  No close  visual companions were
detected by speckle interferometry at SOAR.

{\bf HIP 41214} is a  28-day double-lined  binary. Its  preliminary orbit
presented in GT14 is now definitive. However, the periastron is still
not covered by the observations. Speckle interferometry at SOAR has
not revealed any additional companions.

{\bf HIP 43004} has an orbital  period of 70 days, although the orbit
is still provisional.  No other companions have been found at SOAR.

{\bf HIP  73700}  has a  variable RV. An  orbit
with $P=200$ days is fitted to the data. 

{\bf HIP 79234} is a single-lined binary with a period of 9.07
days, without additional companions.

{\bf HIP 84696} is an acceleration binary for which \citet{Goldin2007}
determined a 4-year  astrometric orbit with an axis  of 23.8\,mas. Our
spectroscopic  period of  3.55 years,  as  well as  other elements  in
common, match that orbit within errors. The astrometric inclination of
$120^\circ$ and  the RV amplitude lead  to the secondary  mass of 0.54
${\cal M}_\odot$.   The semimajor axis  is 70\,mas, and  the estimated
astrometric  axis is 20\,mas,  in agreement  with the  {\it Hipparcos}
orbit. This system is not triple, as far as we know.

{\bf  HIP 88656}  has a  double-lined orbit  with $P=4.27$  days. Five
observations with  CHIRON are  used.  Note the  non-zero eccentricity,
unusual at such short periods.  The ratio of the dip areas corresponds
to the flux ratio of 0.27,  hence the individual $V$ magnitudes of the
components Aa and  Ab are 7.37 and 8.79  mag, respectively. The masses
of the components Aa and  Ab, estimated from their absolute magnitudes
and  the  standard relations,  are  1.25  and  0.82 ${\cal  M}_\odot$,
matching the spectroscopic mass ratio  of 0.71. On the other hand, the
``spectroscopic''  masses  $M  \sin^3  i$  are 0.73  and  0.32  ${\cal
  M}_\odot$ and  imply the orbital  inclination of $ i  =57$\degr. The
widths of the correlation dips  measured with CHIRON correspond to the
approximate  rotational  velocities  $  V  \sin i$  of  13.6  and  7.8
km~s$^{-1}$  for  Aa  and  Ab  that  match  the  expected  synchronous
velocities reasonably  well (a star  of one solar  radius synchronized
and aligned with the orbit would have $ V \sin i = 9.8$ km~s$^{-1}$).

There is  a faint ($V=  13.8$ mag) visual  companion B (A~2595  AB) at
3\farcs0, known  since 1913  and measured for  the last time  in 1972.
Despite  the crowded  sky  in  this area,  the  companion is  physical
because it  keeps a nearly  fixed relative position, while  the proper
motion is moderately  fast.  The projected separation between  A and B
corresponds  to the orbital  period of  the order  of 1.3\,kyr;  no RV
trend  is  noted.  The  star  is  on  the Lick  planet-search  program
\citep{Fischer2014}; however,  the 5  RVs given in  that paper  do not
match the  orbit, possibly because  the double-lined spectrum  was not
accounted for in their data reduction.

{\bf HIP 92140}  has been observed also with the  fiber echelle at the
CTIO 1.5 m mtelescope \citep{LCO}.  We combine the RVM and the echelle
RVs in an orbit with  $P=588$ days. The crosses in Fig.~\ref{fig:orb1}
denote the RVM velocities, the squares are the fiber echelle RVs.  The
orbit can still  be improved, but its long  period requires monitoring
for  several more  years.  This  is a  triple system  with  the visual
companion at 5\farcs8, four  magnitudes fainter than the spectroscopic
pair.  The  orbital period  of the  visual binary is  of the  order of
4\,kyr.

{\bf HIP 104514} has  a slow RV variation with a 266  day period and a
small  amplitude.   The  tertiary   companion  at  3\farcs3  has  been
discovered with Robo-AO \citep{RoboAO}  and later shown to be physical
(co-moving) by \citet{Roberts2015}.

{\bf HIP 112222} is a  bright {\it Hipparcos} acceleration binary with
slow RV variability.   Monitoring during 5 years has  covered one full
orbital  cycle  of 1634$\pm$20  days,  allowing  us  to determine  the
preliminary   elements.    The   minimum   secondary  mass   is   0.34
solar. The semimajor axis is 77\,mas, while the predicted astrometric axis
is 15\,mas. No astrometric orbit has been derived so far, but this
should be an easy task for {\it Gaia}. 

No   additional  companions   to  this   star  are   known.   However,
\citet{SO2011} noted that the K2V  star HIP~112354 has a common proper
motion and distance.  Its  RV of 1.9\,km~s$^{-1}$ also roughly matches
tha centre-of-mass velocity  of our binary, $-2.9$\,km~s$^{-1}$.  This
CPM companion is itself a visual binary BU~711 with a known orbit ($P=
800$  yr).  The  projected  distance  between those  two  binaries  is
0.4\,pc, too  large for a  gravitationally bound pair.   However, they
could be a young co-moving pair of binaries with a common origin.

Among the 19 stars of this Section, 8 have additional components (they
are at least triple).  Partly the large fraction of hierarchies can be
explained  by  the  efforts   of  observers  to  look  for  additional
companions    around     spectroscopic    binaries    within    67\,pc
\citep[e.g.][]{RoboAO}.

%%--------------------------------------------------------------
\section{The triple system HIP 17895}
\label{sec:triple}

\begin{figure}
\centerline{
\includegraphics[width=7cm]{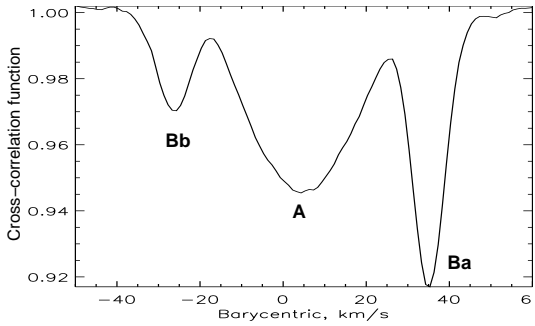}
}
\caption{Cross-correlation function of HIP 17895 recorded with CHIRON on JD 2457257.
\label{fig:CCF} }
\end{figure}

HIP  17895 is  a nearby  solar-type dwarf  with the  HIP2  parallax of
19.63$\pm$0.66 mas.   Double lines  were detected by  the GCS  and the
mass ratio of  0.577$\pm$0.109 was estimated.  At the  same time, this
is  a  visual  binary  YR~23  (WDS J03496$-$0220),  first  resolved  by
\citet{Horch2002} in  2000.76 at  0\farcs28 separation. Such  a binary
would normally be  resolved by {\it Hipparcos}, but  its duplicity has
not  been recognized  by  that experiment.   However, {\it  Hipparcos}
identified  the object  as an  astrometric binary.   It was  not known
whether  the  double lines  were  produced  by the  visual/astrometric
binary itself or whether the system is triple.

The object has been observed with the CHIRON spectrograph at the 1.5-m
telescope at CTIO in the second  half of 2015 and in 2017. The spectra
and  the   RV  measurement  by  cross-correlation   are  described  in
\citep{CHI,Tok2016}.   The  first spectrum  has  clearly shown  triple
lines  (Fig.~\ref{fig:CCF}).   The  broad  central  feature  with  the
largest equivalent  width belongs to  the visual primary  component A,
the two flanking narrow dips correspond to the spectroscopic subsystem
Ba,Bb.  The narrow lines moved on  the time scale of months, while the
central line remained  stationary.  The CHIRON data cover  145 days in
2015, slightly more than half of the orbital cycle.

\begin{figure}
\centerline{
\includegraphics[width=8.5cm]{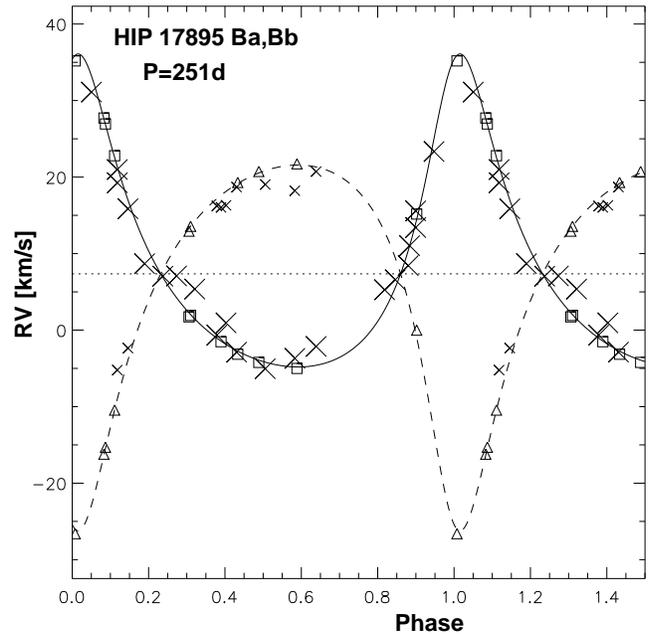}
}
\caption{The RV curve of  HIP 17895 Ba,Bb. The squares and triangles
  stand for the RVs of the primary (Ba) and secondary (Bb) components,
  respectively,  measured with  CHIRON.  The  large and  small crosses
  correspond  to   the  RVM  data  of  lower   accuracy.
\label{fig:orb3} }
\end{figure}

We placed  this star  on the RVM  observing programme in  2012.  These
observations  are used  in the  orbit calculation  to extend  the time
coverage  and  constrain  the  orbital  period.   However,  the  lower
spectral  resolution  of  the  RVM,  its lower  sensitivity,  and  the
triple-lined nature of  the spectrum all lead to  the relatively large
errors of the  RVM velocities. The elements of  the double-lined orbit
of Ba,Bb  are given in Table~\ref{tab:orb},  the RV curve  is given in
Fig.~\ref{fig:orb3}. In the final least-squares adjustment, the errors
of  the CHIRON RVs  (and the  corresponding weights)  were set  to 0.3
km~s$^{-1}$,  while   the  errors  of   the  RVM  velocities   were  increased
artificially to reflect their lower accuracy.

Although the measurements  of the visual pair A,B  in 2001--2014 cover
only  a small  fraction of  its orbit,  its preliminary  elements with
$P=54$  yr were  determined  by \citet{RoboAO}.   Here  this orbit  is
updated by  using the latest data,  a different set of  weights, and a
fixed eccentricity of 0.6.  As the orbit is poorly constrained anyway,
this is  acceptable.  A larger eccentricity corresponds  to the larger
mass sum.  By choosing $e=0.6$,  we get the  expected mass sum  of 2.9
${\cal  M}_\odot$.  The  orbit implies  a separation  of  0\farcs22 in
1991.25, so  the non-resolution by {\it Hipparcos}  is still difficult
to explain.  If we adopt a shorter period (meaning a closer separation
in 1991), the mass sum becomes too large (e.g. 11 ${\cal M}_\odot$ for
$P=40$ yr).  The new orbital elements  of A,B are: $P=51.1$ yr, $T_0 =
1988.3$,  $e=0.6$,  $a=0\farcs383$, $\Omega_A  =  68\fdg6$, $\omega  =
286\fdg5$, $i=121\fdg1$.

The visual  elements and  the expected mass  sum correspond to  the RV
amplitudes in  the outer  orbit of  $K_A + K_B  = 14$  km~s$^{-1}$ and
match  the  RV difference  of  $\sim$2\,km~s$^{-1}$  between  A and  B
measured  in  2015 (the  RVs  of  A  measured from  the  well-resolved
profiles are 4.8, 5.4, and 5.0 km~s$^{-1}$ on JD 2457257, 2457276, and
2457277, respectively). This means that the true ascending node of the
visual orbit is already known.  Small orbital coverage prevents us
from  measuring the masses and orbital parallaxe from the combined
visual/spectroscopic orbit of the outer pair A,B.  Large variations of
the  RV are  predicted to  occur  at the  time of  the next  periastron
passage around 2040.

Individual  magnitudes  and masses  of  all  three  components can  be
deduced from  the available data.  The speckle  interferometry at SOAR
yields the relative photometry of the  A,B pair: $\Delta y = 0.78$ mag
(two measures)  and $\Delta I =  0.57$ mag (one  measure).  The latter
matches  $\Delta I  = 0.56$  mag given  by \citet{Horch2002}.   On the
other  hand, the  equivalent width  of the  individual details  in the
cross-correlation  (Fig.~\ref{fig:CCF})   is  0.90,  0.32,   and  0.11
km~s$^{-1}$ for A, Ba, and Bb, respectively.  The flux ratio between A
and B  is thus 0.48$\pm$0.05 or  0.80 mag, in good  agreement with the
speckle photometry. So, the individual $V$ magnitudes of A, Ba, and Bb
are estimated to be 7.67,  8.74, and 9.92 mag, respectively.  Standard
relations  for main sequence  stars and  the {\it  Hipparcos} parallax
lead to  the masses of 1.14,  0.96, and 0.80 ${\cal  M}_\odot$, or the
mass  sum of  2.9 ${\cal  M}_\odot$.  These  estimates agree  with the
directly measured mass  ratio in the inner binary,  $q_{\rm Ba,Bb} =
0.85$.  All  three stars  in this triple  system apparently  match the
standard main-sequence relation between mass and absolute magnitude.

The spectroscopic mass sum in  the inner subsystem $({\cal M}_{\rm Ba}
+ {\cal  M}_{\rm Ba}) \sin^3 i  = 1.72 {\cal  M}_\odot$ is just slightly
less than the estimated mass sum of $1.76 {\cal M}_\odot$.  This means
that the orbit of Ba,Bb  has a high inclination $i_{\rm Ba,Bb} \approx
80^\circ$ or  $i_{\rm Ba,Bb}  \approx 100^\circ$.  The  inclination of
the  outer orbit is  $i_{\rm A,B}=121$\degr.  Therefore, the  inner and
outer orbits possibly have  small mutual inclination.

The  semimajor axis  of  the  inner orbit  is  18.5\,mas. Knowing  the
inclination, we can  compute the separation between Ba  and Bb. At the
moment of the  SOAR observation in 2011.04 it  was 15\,mas, just below
the diffraction limit. There is a  hint that the inner pair was indeed
partially  resolved, but  no reliable  measure can  be  extracted. The
maximum separation of about 26\,mas and $\Delta V = 1.2$ mag mean that
the subsystem Ba,Bb can be resolved  at 4-m telescopes and certainly at 8-m
ones and with long-baseline interferometers like CHARA. Several resolved
observation  of  Ba,Bb  would   establish  its visual orbit  and  the  mutual
inclination  in  this  triple  system.   The  250-day  motion  of  the
photo-centre caused by the subsystem  is large enough to be detectable
by {\it Gaia} or by the residuals in the outer orbit.

The  spectral lines  of the  component A  are broadened  by  the axial
rotation of  $V \sin i  \approx 24$ km~s$^{-1}$   corresponding to
  the rotational period of  $<$2.6 days. According to the age-rotation
  relation proposed by \citet{Barnes2007},  the age might be less than
  200\,Myr.  The large  $V \sin i$  also  suggests a high inclination
of the  rotation axis,  matching in this  respect the  highly inclined
orbits.  Coronal X-ray emission from this multiple system was detected
by {\it ROSAT} (RX J0349.6$-$0219).

%%--------------------------------------------------------------
\section{Stars without orbits}
\label{sec:var}

\begin{figure*}
\centerline{
\includegraphics[width=16cm]{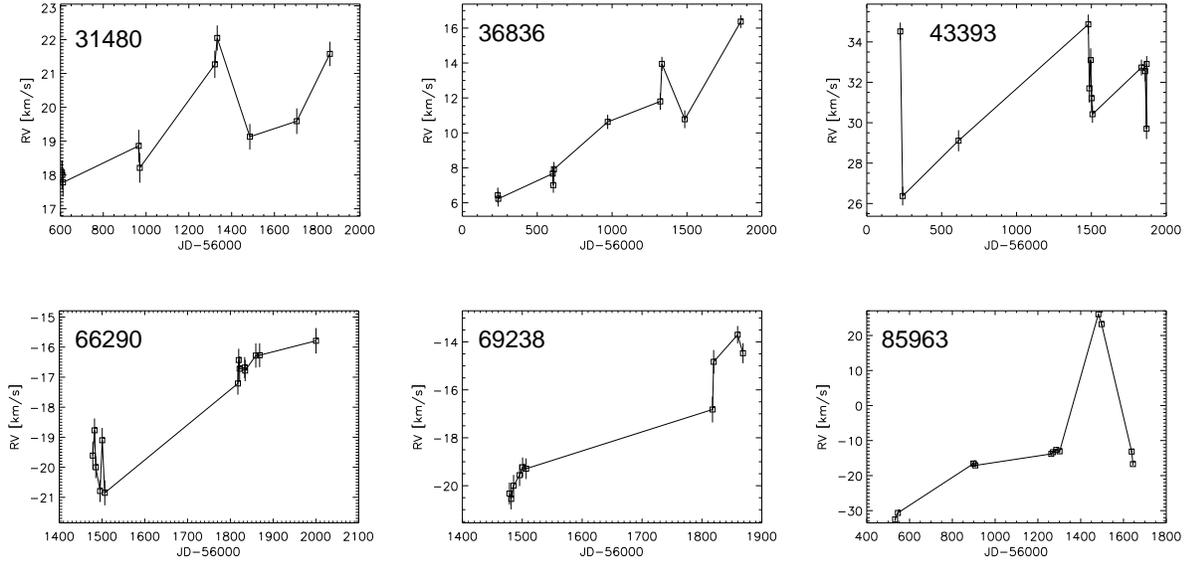}
}
\caption{Plots of RV vs. time for six stars with slow RV variation. The
  {\it Hipparcos} numbers are indicated in the plots.
\label{fig:trend} }
\end{figure*}

\begin{table}
\centering
\caption{Average RVs of stars without orbits}
\label{tab:RVplus}
\medskip
\begin{tabular}{c c rrr  } 
%\hline
\hline
HIP  & $N$ &  $\langle$RV$\rangle$  & $\sigma$ & $\Delta T$ \\
     &     & km~s$^{-1}$ & km~s$^{-1}$ &  days \\
\hline
 20693 & 27 &  31.51   & 1.66  &   1483 \\ 
 24076 & 22 &   6.87   & 1.14  &   1482 \\
 26444 & 11 &  47.41   & 0.85  &   1102 \\
 27878 & 19 &  32.70   & 1.04  &   1481 \\
 31480 & 10 &  19.45   & 1.61  &   1252 \\
 36836 & 10 &   9.88   & 3.43  &   1622 \\
 43393 & 12 &  31.60   & 2.41  &   1645 \\
 48391 & 15 & $-$12.09 & 1.04  &    390 \\
 55982 & 14 &   8.68   & 4.94  &    389 \\
 56282 & 12 &   9.53   & 6.59  &    388 \\
 66290 & 14 & $-$17.95 & 1.82  &    521 \\
 69238 & 10 & $-$17.88 & 2.66  &    389 \\
 73765 & 23 & $-$12.10 & 9.20  &   1341 \\
 81312 & 33 &  $-$5.31 & 0.50  &   1128 \\
% 84696 & 13 & $-$18.61 & 3.71  &   1111 \\
 85963 & 13 & $-$11.29 &17.20  &   1111 \\
109122 & 61 &  $-$6.10 & 6.31  &   1433 \\
%112222 & 31 &  $-$1.44 & 2.65  &   1482 \\
\hline
%\hline
\end{tabular}
\end{table}

Table~\ref{tab:RVplus}  reports  on  the  16 stars  observed  in  this
programme with  RVM but lacking orbits.  It gives the  total number of
RVs  $N$,  the average  RV,  the  unweighted  r.m.s.  scatter  of  the
velocities  $\sigma$, and  the  total time  span  of the  observations
$\Delta  T$  in   days.  The  individual  RVs  are   provided  in  the
Supplementary material, Table~5. 

According   to   the   GCS,   all  those   stars   are   spectroscopic
binaries. However, some of them turned  out to have a constant RV (HIP
20693,  26444,  27878,  48391,  81312).   The  spurious  detection  of
variability in the  GCS could have been caused  by an outlying measure
or by pointing  a wrong star. Another group of stars  show only a slow
RV variation (e.g.  linear trends) indicative of long orbital periods;
for some  of those  an astrometric acceleration  was also  detected by
{\it Hipparcos}  (HIP 31480, 36836, 43393, 66290,  69238, 85963). 
  The  plots   of  RV  vs.   time   for  these  stars   are  given  in
  Fig.~\ref{fig:trend}.     The  remaining objects  are spectroscopic
binaries  for which  our  data  do not  yet  allow orbit  calculation.
Individual comments are provided below.  All periods proposed below
  are tentative.

{\bf HIP 20693}  has wide lines that could explain  the scatter of the
RVs.   A  period of  2.458  days can  be  found with  an
amplitude of 1.8 km~s$^{-1}$, but this ``orbit'' is not convincing. 

{\bf  HIP  24076}  is  the  visual  binary  A~484  with  nearly  equal
components.   The  RV  is  likely  constant, while  the  double  lines
detected by the GCS could arise at periastron in the eccentric 19-year
orbit \citep{Tok2017}.  The RVs favor this eccentric orbit, instead of
the previously computed 37.6-year circular orbit.

{\bf HIP 26444  } apparently has a constant RV  of 47.4 km~s$^{-1}$, matching
the RV  of 47.7  km~s$^{-1}$ measured with  CHIRON. The  GCS gives the  RV of
49.8~km~s$^{-1}$.

{\bf  HIP 43393}  has a  slow  RV variation.  The 11  RVs measured  by
\citet{Latham2002} have the average value of 33.22 km~s$^{-1}$ and the
rms  scatter  of  0.64  km~s$^{-1}$;  we  measure  the  average  RV  of
31.6~km~s$^{-1}$. 

{\bf HIP 55982} might have an orbital period of 35 days. 

{\bf HIP 56282} can have an orbital period of 121.7 or 72.4 days. There
is a CPM companion at 15\farcs8. 

{\bf  HIP 73765}  has  a large  RV  variation that  can correspond  to
periods of  1.48 or 2.97  days. The coverage  of the RV curve  is poor.
  This is an X-ray source. 

{\bf  HIP  81312}  is  a   metal-poor  star  with  a  constant  RV  of
$-5.3$~km~s$^{-1}$, in agreement  with \citet{Latham2002} who measured
the average  RV of  $-5.06$ km~s$^{-1}$ with  the rms scatter  of 0.76
km~s$^{-1}$.   Yet,   the  GCS  announced  the  RV   variation  by  39
km~s$^{-1}$.

{\bf HIP  85963} has a variable  RV and is an  acceleration binary. It
was tentatively resolved  at SOAR in 2014 into  a close 0\farcs09 pair
TOK~417 with an  estimated period of 10 years.  However, the resolution
was not confirmed in 2015 and 2017. The large RV variation may
imply a very eccentric orbit.

{\bf  HIP 109122}  is  the {\it  Hipparcos}  acceleration binary  with
variable RV  (GCS).  We confirm  the RV variability. A  crude 3.2-year
orbit   can  be fitted  to  the  RVs, but  more
observations are  necessary before it  can be published. The  star has
not been  resolved by  speckle interferometry, presumably  because the
secondary is too faint (the estimated semimajor axis is $\sim$50\,mas).

\section*{Acknowledgments}

This work was  supported by the Russian Foundation  for Basic Research
(project  code 14-02-00472) and  the grant  of the  Russian Scientific
Foundation No. 14-22-00041.  We thank the administration of the Simeiz
Section  of  the  Crimean  Astrophysical  Observatory  for  allocating
observing time  on the  1-m telescope. Some  observations made  at the
CTIO  1.5-m telescope  were  used.   Constructive  comments by  the
  anonymous Referee are gratefully acknowledged.

%%--------------------------------------------------------------
%\section{}
%\label{sec:}

%%--------------------------------------------------------------
%\section{}
%\label{sec:}

%%--------------------------------------------------------------
%\section{}
%\label{sec:}

\section{Supplementary material}

Table~4  gives the  individual RVs  and residuals  to the  orbits. Its
columns contain  the {\it  Hipparcos} number, the  heliocentric Julian
date of observation minus 2400000, the RV, its error, and the residual
to the orbit, all in km~s$^{-1}$.  The last column of Table~4 contains
the  flag denoting  the component  (a for  the primary  and b  for the
secondary). Letters  F or C are  added to the component  flag to mark
the  RVs measured  with the  fiber echelle  and  CHIRON, respectively.
Table~5 gives the individual RVs  of stars without orbits, in the same
form  except that  there  are no  columns  with the  residuals and  the
component flag.  All RVs listed in Table~5 are measured with the
  RVM. 

\label{lastpage}

\end{document}